\documentclass{emulateapj}
\usepackage{apjfonts}
\usepackage{rotating}
\usepackage{multirow}

\newcommand{\pivec}{\mbox{\boldmath $\pi$}}

\lefthead{PARK ET AL.}
\righthead{ }

\begin{document}
\title{OGLE-2013-BLG-0578L: MICROLENSING BINARY COMPOSED OF A BROWN DWARF AND AN M DWARF} 

\author{
H. Park$^{1}$, 
A. Udalski$^{2,5}$,
C. Han$^{1,6}$, \\
AND \\
R. Poleski$^{2,3}$,
J. Skowron$^{2}$, 
S. Koz{\l}owski$^{2}$, 
{\L}. Wyrzykowski$^{2,4}$,
M. \,K. Szyma\'nski$^{2}$, 
P. Pietrukowicz$^{2}$, 
G. Pietrzy\'nski$^{2}$, \\
I. Soszy\'nski$^{2}$, 
K. Ulaczyk$^{2}$ \\ 
(The OGLE Collaboration)
}

\bigskip\bigskip
\affil{$^{1}$Department of Physics, Institute for Astrophysics, Chungbuk National University, Cheongju 371-763, Korea}
\affil{$^{2}$Warsaw University Observatory, Al. Ujazdowskie 4, 00-478 Warszawa, Poland}
\affil{$^{3}$Department of Astronomy, Ohio State University, 140 West 18th Avenue, Columbus, OH 43210, USA}
\affil{$^{4}$Institute of Astronomy, University of Cambridge, Madingley Road, Cambridge CB3 0HA, UK}

\footnotetext[5]{
The OGLE Collaboration}
\footnotetext[6]{
Corresponding author}

\begin{abstract}
Determining physical parameters of binary microlenses is hampered by the lack of information 
about the angular Einstein radius due to the difficulty of resolving caustic crossings. In 
this paper, we present the analysis of the binary microlensing event OGLE-2013-BLG-0578, for 
which the caustic exit was precisely predicted in advance from real-time analysis, enabling 
to densely resolve the caustic crossing and to measure the Einstein radius. From the mass 
measurement of the lens system based on the Einstein radius combined with the additional 
information about the lens parallax, we identify that the lens is a binary that is composed 
of a late-type M-dwarf primary and a substellar brown-dwarf companion. The event demonstrates 
the capability of current real-time microlensing modeling and the usefulness of microlensing 
in detecting and characterizing faint or dark objects in the Galaxy. 
\end{abstract}


\section{INTRODUCTION}

It is known that low-mass stars comprise a significant fraction of stars in the Solar neighborhood 
and the Galaxy as a whole. The Galaxy may be teeming with even smaller mass brown dwarfs. Therefore, 
studying the abundance and properties of low-mass stars and brown dwarfs is of fundamental importance. 
There have been surveys searching for very low-mass (VLM) objects \citep{reid08,aberasturi14}, but 
these surveys are limited to the immediate solar neighborhood. As a result, the sample of VLM objects 
is small despite their intrinsic numerosity and thus our understanding about VLM objects is poor. 

Microlensing surveys detect objects through their gravitational fields rather than their radiation 
and thus microlensing can provide a powerful probe of VLM objects. However, the weakness of microlensing 
is that it is difficult to determine the lens mass for general microlensing events. This difficulty 
arises due to the fact that the time scale of an event, which is the only observable related to the 
physical lens parameters, results from the combination of the lens mass, distance and the relative 
lens-source transverse speed. As a result, it is difficult to identify and characterize VLM objects 
although a significant fraction of lensing events are believed to be produced by these objects. 

However, it is possible to uniquely determine the physical lens parameters and thus identify VLM 
objects for a subset of lensing events produced by lenses composed of two masses. For unique 
determinations of the physical lens parameters, it is required to simultaneously measure the 
angular Einstein radius $\theta_{\rm E}$ and the microlens parallax $\pi_{\rm E}$ that are related 
to the lens mass $M$ and distance to the lens $D_{\rm L}$ by
\begin{equation}
M_{\rm tot}={\theta_{\rm E} \over \kappa\pi_{\rm E}};\qquad  
D_{\rm L}={{\rm AU}\over \pi_{\rm E}\theta_{\rm E}+\pi_{\rm S}},
\label{eq1}
\end{equation}
respectively \citep{gould00}. Here $\kappa=4G/(c^{2}{\rm AU})$, $\pi_{\rm S}={\rm AU}/D_{\rm S}$ 
is the parallax of the source star, and $D_{\rm S}$ is the distance to the source. The angular 
Einstein radius is estimated by analyzing deviations in lensing light curves caused by the finite 
size of the lensed source stars \citep{gould94,nemiroff94,witt94}. For a single-lens events, 
finite-source effects can be detected only for a very small fraction of extremely high-magnification 
events where the lens-source separation at the peak magnification is equivalent to the source size. 
On the other hand, light curves of binary-lens events usually result from caustic crossings during 
which finite-source effects become important and thus the chance to detect finite-source effects 
and measuring the Einstein radius is high. Furthermore, binary-lens events tend to have longer 
time scales than single-lens events and this also contributes to the higher chance to measure the 
lens parallax. In fact, most known VLM lensing objects were identified through the channel of 
binary-lens events \citep{hwang10,shin12,choi13,han13,park13,jung15}. 

Despite the usefulness of binary-lens events, the chance to identify VLM objects has been low. 
One important reason for the low chance is that caustic crossings last for very short periods of 
time. The duration of a caustic crossing is 
\begin{equation}
t_{\rm cc} = {2\rho \over \sin\phi}t_{\rm E},
\label{eq2}
\end{equation}
where $t_{\rm E}$ is the Einstein time scale of the event, $\phi$ is the entrance angle of the 
source star with respect to the caustic line, and $\rho=\theta_{\ast}/\theta_{\rm E}$ is the 
angular source radius $\theta_{\ast}$ normalized to the angular Einstein radius $\theta_{\rm E}$. 
Considering that the Einstein time scale is $\sim({\cal O}) 10$ days and the Einstein radius is 
$\sim({\cal O})$ milli-arcsec 
for typical Galactic microlensing events, the duration of a caustic crossing is $\sim({\cal O})$ hours for 
Galactic bulge source stars with angular stellar radii of $\sim({\cal O})$ 1--10 $\mu$-arcsec. Therefore, 
it is difficult to densely resolve caustic crossings from surveys that are being carried out with 
over hourly observational cadences. 

Another reason for the low chance of resolving caustic crossings is the difficulty in predicting 
their occurrence. Caustics produced by a binary lens form a single or multiple sets of closed curves 
and thus caustic crossings always come in pairs. Although it is difficult to predict the first 
crossing (caustic entrance) based on the fraction of the light curve before the caustic entrance, 
the second crossing (caustic exit) is guaranteed after the caustic entrance. To resolve the short-lasting 
caustic exit, it is required to precisely predict the time of the caustic crossing so that observation 
can be focused to resolve caustic crossings. This requires vigilant modeling of a lensing event conducted 
with the progress of the event followed by intensive follow-up observation. 

In this paper, we report the discovery of a VLM binary that was detected from the caustic-crossing 
binary-lens microlensing event OGLE-2013-BLG-0578. The caustic exit of the event was precisely predicted 
by real-time modeling, enabling dense resolution and complete coverage of the caustic crossing. Combined 
with the Einstein radius measured from the caustic-crossing part of the light curve and the lens 
parallax measured from the long-term deviation induced by the Earth's parallactic motion, we uniquely 
measure the lens mass and identify that the lens is a VLM binary composed of a low-mass star and 
a brown dwarf.

\section{Observation}

The microlensing event OGLE-2012-BLG-0578 occurred on a star located toward the Galactic Bulge direction. 
The equatorial coordinates of the lensed star are $(\alpha,\delta)_{\rm J2000}=(17^{\rm h}59^{\rm m}
59^{\rm s}\hskip-2pt.85, -29^{\circ}44'06''\hskip-2pt.9)$, that correspond to the Galactic coordinates 
$(l,b)=(0^{\circ}\hskip-2pt.90, -3^{\circ}\hskip-2pt.10)$. The event was first noticed on April 22, 
2013 from survey observations conducted by the Optical Gravitational Lensing Experiment 
\citep[OGLE:][]{udalski03} using the 1.3m Warsaw telescope at the Las Campanas Observatory in Chile. 
Images were taken using primarily in {\it I}-band filter and some {\it V}-band images were also taken 
to constrain the lensed star (source).

\begin{figure}[ht]
\epsscale{1.15}
\plotone{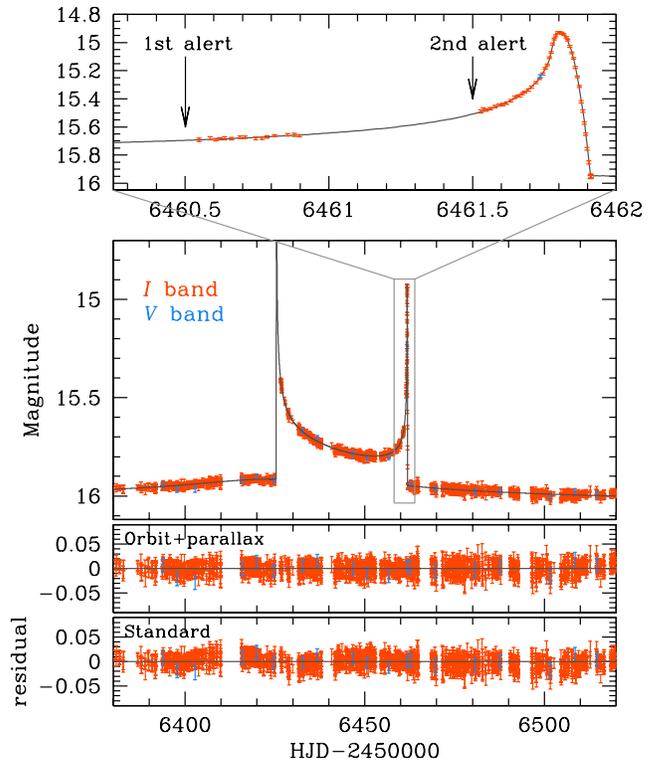}
\caption{\label{fig:one}
Light curve of OGLE-2013-BLG-0578. Also plotted is the best-fit model 
(``orbit+parallax'' model with $u_{0}>0$). The two bottom panels show 
the residuals from the modeling with and without considering 
lens-orbital and parallax effects. The upper panel shows the enlargement 
of the caustic-exit region and the times of caustic-crossing alerts.
}\end{figure}

In Figure~\ref{fig:one}, we present the light curve of the event. The light curve shows two distinctive 
spikes that are characteristic features of a caustic-crossing binary-lens event. The caustic crossings 
occurred at ${\rm HJD}'={\rm HJD}-2450000\sim6426.0$ and $6461.8$. Although the event was first noticed 
before the caustic crossings, the caustic entrance was missed. With the progress of the event, it became 
clear that the event was produced by a binary lens from the characteristic ``U''-shape trough between the 
caustic crossings.

Although the first caustic crossing was missed, the second crossing was densely resolved. Resolving the 
crossing became possible with the prediction of the caustic crossing from vigilant modeling of the light 
curve followed by intensive follow-up observation. It is known that reliable prediction of the second 
caustic crossing is difficult based on the light curve before the minimum between the two caustic crossings 
\citep{jaro01}. With the emergence of the correct model after passing the caustic trough, we focused on 
the prediction of the exact caustic-crossing time. The first alert of the caustic exit was issued on June 17, 2013, 
1.3 days before the actual caustic exit. On the next day, the second alert was issued to predict more 
refined time of the caustic exit. In response to the alert, the OGLE experiment, which is usually operated 
in survey mode, entered into ``following-up'' mode observation by increasing the observation cadence. 
Thanks to the intensive follow-up observation, the caustic exit was completely and densely covered. See the 
upper panel of Figure~\ref{fig:one}. 

In order to securely measure the baseline magnitude and detect possible higher-order effects, observation 
was continued after the caustic exit to the end of the Bulge season. From these observations, we obtain 
2284 and 27 images taken in {\it I} and {\it V} bands, respectively. Photometry of the event was done by 
using the customized pipeline \citep{udalski03} that is based on the Difference Imaging Analysis method 
\citep{alard98,wozniak00}. We note that the {\it I}-band data are used for light curve analysis, while 
the {\it V}-band data are used for the investigation of the source type. 

It is known that photometric errors estimated by an automatic pipeline are often underestimated and thus 
errors should be readjusted. We readjust error bars by
\begin{equation}
e'=k(e^{2}+e_{\rm min}^{2})^{1/2}.
\label{eq3}
\end{equation}
Here $e_{\rm min}$ is a term used to make the cumulative distribution function of $\chi^{2}$ as a function 
of lensing magnification becomes linear. This process is needed to ensure that the dispersion of data points 
is consistent with error bars of the source brightness. The other term $k$ is a scaling factor used to make 
$\chi^{2}$ per degree of freedom (dof) becomes unity.

\section{Analysis}

We analyze the event by searching for the set of lensing parameters (lensing solution) that best describe 
the observed light curve. Basic description of a binary-lens event requires seven standard lensing parameters. 
Three of these parameters describe the source-lens approach including the time of the closest approach of the 
source to a reference position of the lens, $t_{0}$, the lens-source separation at $t_{0}$ in units of the 
Einstein radius, $u_{0}$, and the time required for the source to cross the Einstein radius, $t_{\rm E}$ (Einstein 
time scale). In our analysis, we set the center of the mass of the binary lens as a reference position in the 
lens plane. Another two parameters describe the binary lens including the projected binary separation in units 
of the Einstein radius, $s$ (normalized separation), and the mass ratio between the lens components, $q$. Due 
to the asymmetry of the gravitational field around the binary lens, it is needed to define the angle between 
the source trajectory and the binary axis, $\alpha$ (source trajectory angle). The last parameter is the 
normalized source radius $\rho$, which is needed to describe the caustic-crossing parts of the light curve 
that are affected by finite-source effects. 

It is often needed to consider higher-order effects in order to precisely describe lensing light curves and 
this requires to include additional lensing parameters. For long time-scale events, such effects are caused 
by the positional change of an observer induced by the orbital motion of the Earth around the Sun 
\citep[``parallax effect'':][]{gould92} and/or the change of the binary separation and orientation caused by the 
orbital motion of the lens \citep[``lens orbital effect'':][]{dominik98,albrow00}. The analyzed event lasted 
throughout the whole Bulge season and thus these effects can be important. The parallax effect is described 
by two parameters, $\pi_{\rm E,\it N}$ and $\pi_{\rm E,\it E}$, that are the two components of the lens 
parallax vector ${\pivec}_{\rm E}$ projected onto the sky along the north and east equatorial coordinates, 
respectively. The direction of the lens parallax vector corresponds to the relative lens-source proper motion 
and its magnitude corresponds to the relative lens-source parallax $\pi_{\rm rel}={\rm AU}(D_{\rm L}^{-1}
-D_{\rm S}^{-1})$ scaled to the Einstein radius of the lens, i.e.,
\begin{equation}
\pi_{\rm E} = { \pi_{\rm rel} \over \theta_{\rm E} }.
\label{eq4}
\end{equation}
To the first-order approximation, the lens orbital effect is described by two parameters, $ds/dt$ and $d\alpha/dt$, 
which are the change rates of the normalized binary separation and the source trajectory angle, respectively. 

\begin{figure}[ht]
\epsscale{1.15}
\plotone{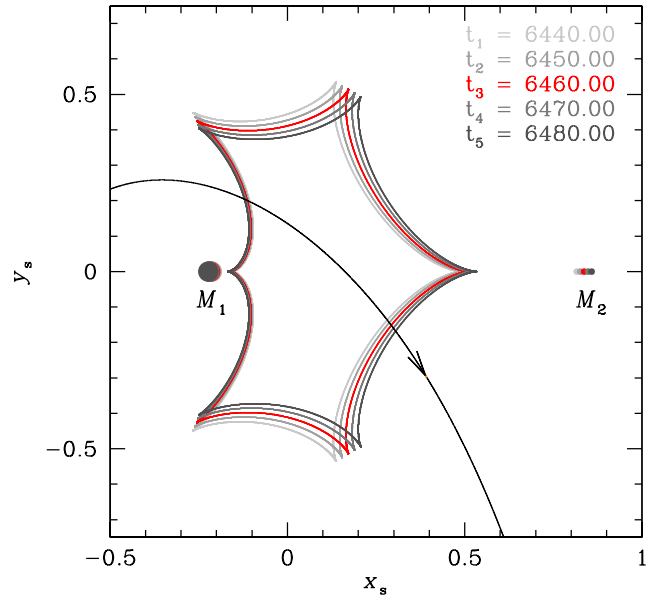}
\caption{\label{fig:two}
The source trajectory (the curve with an arrow) with respect to 
the caustic (the closed curve with 6 cusps). The two small dots 
marked by $M_1$ and $M_2$ represent the positions of the binary 
lens components. All lengths are scaled by the angular Einstein 
radius corresponding to the total mass of the binary lens. Due 
to the change of the relative positions of the binary lens components 
caused by the orbital motion, the caustic varies in time. We present 
4 caustics and lens positions corresponding to the times marked 
in right upper corner of the panel. The source trajectory is curved 
due to the combination of the lens-orbital and parallax effects. 
}\end{figure}

To model caustic-crossing parts of the light curve, it is needed to compute magnifications affected by 
finite-source effects. To compute finite-source magnifications, we use the numerical method of the inverse 
ray-shooting technique \citep{kayser86,schneider86} in the immediate neighboring region around caustics and 
the semi-analytic hexadecapole approximation \citep{pejcha09,gould08} in the outer region surrounding caustics. 
We consider the effects of the surface brightness variation of the source star. The surface brightness is 
modeled as 
\begin{equation}
S_{\lambda}\propto 1-\Gamma_{\lambda} \left(1-{3\over 2}\cos \psi \right),
\label{eq5}
\end{equation}
where $\Gamma_{\lambda}$ is the linear limb-darkening coefficient, $\lambda$ is the passband, and $\psi$ is 
the angle between the line of sight toward the source star and the normal to the source surface. We adopt 
the limb-darkening coefficients $(\Gamma_{V}, \Gamma_{I})=(0.62, 0.45)$ from \citet{claret00} based on the 
source type. The source type is determined based on its de-reddened color and brightness. See Section 4 
for details about how the source type is determined.

Searching for the best-fit solution of the lensing parameters is carried out based on the combination of 
grid-search and downhill approaches. We set $(s,q,\alpha)$ as grid parameters because lensing magnifications 
can vary dramatically with small changes of these parameters. On the other hand, magnifications vary smoothly 
with changes of the remaining parameters, and thus we search for these parameters by using a downhill approach. 
We use the Markov Chain Monte Carlo (MCMC) method for the downhill approach. Searching for solutions throughout 
the grid-parameter spaces is important because it enables one to check the possible existence of degenerate 
solutions where different combinations of the lensing parameters result in similar light curves.

In the initial search for solutions, we conduct modeling of the light curve based on the 7 basic binary-lensing 
parameters (``standard model''). From this, it is found that the event was produced by a binary with a projected 
separation very close to the Einstein radius, i.e. $s\sim1.0$. Caustics for such a resonant binary form a single 
big closed curve with 6 cusps. The two spikes of the light curve were produced by the source trajectory passing 
diagonally through the caustic. See Figure~\ref{fig:two} where we present the source trajectory with respect to 
the caustic\footnotemark[7]. The estimated mass ratio between the binary components is $q\sim0.2-0.3$.
\footnotetext[7]{
We note that the source trajectory is curved due to the combination of 
the lens-orbital and parallax effects. We also note that the caustic 
varies in time due to the positional change of the binary-lens components 
caused by the orbital motion.} 

\begin{figure}[ht]
\epsscale{1.15}
\plotone{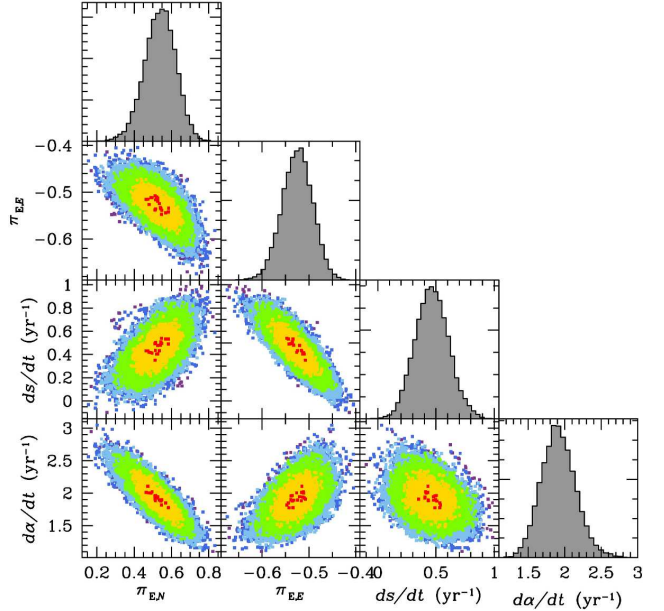}
\caption{\label{fig:three}
Contours of $\Delta\chi^{2}$ in the space of the parallax and 
lens-orbital parameters for the best-fit model ($u_{0}>0$ 
model). Contours marked in different colors represent the 
regions with $\Delta\chi^{2}<1$ (red), $4$ (yellow), $9$ 
(green), $16$ (sky blue), $25$ (blue), and $36$ (purple). 
}\end{figure}

Although the standard model provides a fit that matches the overall pattern of the observed light curve, it is 
found that there exists some residual in the region around ${\rm HJD}'\sim6410$. See the bottom panel of 
Figure~\ref{fig:one}. We, therefore, check whether higher-order effects improve fit. We find that separate 
consideration of the parallax effect (``parallax model'') and the lens-orbital effect (``orbital model'') 
improves fit by $\Delta\chi^{2}=579$ and $344$, respectively. When both effects are simultaneously considered 
(``orbit+parallax model''), the fit improves by $\Delta\chi^{2}=616$, implying that both effects are important. 
In Figure~\ref{fig:three}, we present the contours of $\Delta\chi^{2}$ in the space of the higher-order lensing 
parameters. Contours marked in different colors represent the regions with $\Delta\chi^{2}<1$ (red), 
$4$ (yellow), $9$ (green), $16$ (sky blue), $25$ (blue), and $36$ (purple). It shows that the higher-order 
effects are clearly detected. Considering the time gap between the caustic crossings that is approximately 
a month and the long duration of the event that lasted throughout the whole Bulge season, the importance of 
the higher-order effects is somewhat expected. 

\begin{deluxetable}{lrr}
\tablecaption{Lensing parameters \label{table:one}}
\tablewidth{0pt}
\tablehead{
\multicolumn{1}{c}{Parameter} &
\multicolumn{1}{c}{$u_{0}>0$}  &
\multicolumn{1}{c}{$u_{0}<0$}  
}
\startdata
$\chi^2/{\rm dof}$                         & 2303.1/2300           & 2328.1/2300           \\
$t_{0}$ (HJD)                              & 2456440.13 $\pm$ 0.11 & 2456439.99 $\pm$ 0.08 \\
$u_{0}$                                    & 0.109 $\pm$ 0.004     & -0.120 $\pm$ 0.003    \\
$t_{\rm E}$ (days)                         & 72.11 $\pm$ 0.84      & 70.79 $\pm$ 0.10      \\
$s$                                        & 1.027 $\pm$ 0.004     & 1.035 $\pm$ 0.002     \\
$q$                                        & 0.260 $\pm$ 0.004     & 0.241 $\pm$ 0.003     \\
$\alpha$ (rad)                             & 0.676 $\pm$ 0.007     & -0.647 $\pm$ 0.007    \\
$\rho$ ($10^{-3}$)                         & 0.96 $\pm$ 0.01       & 0.97 $\pm$ 0.01       \\
$\pi_{\rm E,\it N}$                        & 0.54 $\pm$ 0.08       & -0.51 $\pm$ 0.05      \\
$\pi_{\rm E,\it E}$                        & -0.53 $\pm$ 0.03      & -0.72 $\pm$ 0.02      \\
$ds/dt$ (${\rm yr}^{-1}$)                  & 0.48 $\pm$ 0.14       & 0.40 $\pm$ 0.03       \\
$d\alpha/dt$ (${\rm rad}$ ${\rm yr}^{-1}$) & 1.89 $\pm$ 0.24       & -1.75 $\pm$ 0.13
\enddata                                              
\end{deluxetable}

It is known that lensing events with higher-order effects are subject to the degeneracy caused by the mirror 
symmetry of the source trajectory with respect to the binary axis \citep{skowron11}. This so-called ``ecliptic 
degeneracy'' is important for Galactic Bulge events that occur near the ecliptic plane. The pair of the solutions 
resulting from this degeneracy have almost identical parameters except $(u_{0}, \alpha, \pi_{\rm E,\it N}, d\alpha/dt)
\rightarrow -(u_0,\alpha,\pi_{{\rm E}, N},d\alpha/dt)$. It is found that $u_{0}>0$ is marginally preferred 
over the $u_{0}<0$ solution by $\Delta\chi^{2}=25.0$, which corresponds to formally $\sim5\sigma$ level difference. 
However, this level of $\Delta\chi^{2}$ can often occur due to systematics in data and thus one cannot 
completely rule out the $u_{0}<0$ solution. In Table~\ref{table:one}, we present the best-fit parameters of both 
$u_{0}>0$ and $u_{0}<0$ solutions. We note that the uncertainties of the lensing parameters are estimated based 
on the distributions of the parameters obtained from the MCMC chain of the solution. We also present the best-fit 
model light curve ($u_{0}>0$ solution) and the corresponding source trajectory with respect to the lens and 
caustic in Figures~\ref{fig:one} and~\ref{fig:two}, respectively. 

\begin{figure}[ht]
\epsscale{1.15}
\plotone{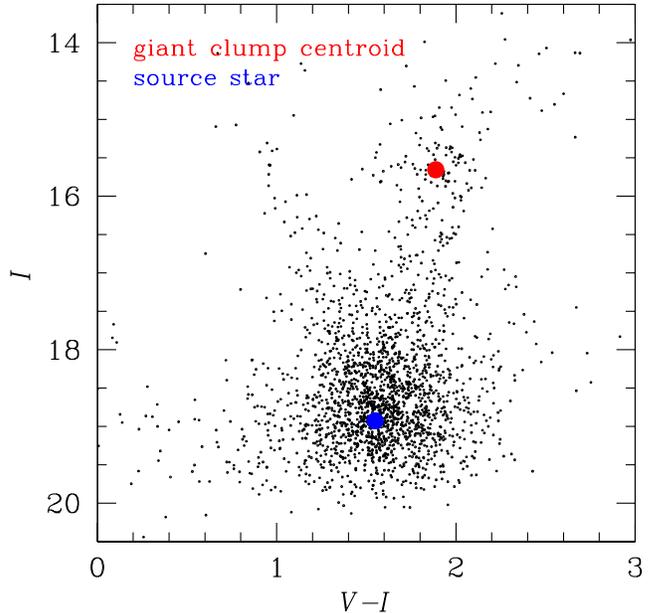}
\caption{\label{fig:four}
Location of the lensed star with respect to the centroid of giant 
clump in the color-magnitude diagram of neighboring stars.
}\end{figure}

\section{Physical Parameters}

By detecting both finite-source and parallax effects, one can measure the angular Einstein radius and the lens 
parallax, which are the two quantities needed to determine the mass and distance to the lens. The lens parallax 
is estimated by $\pi_{\rm E}=(\pi_{\rm E,\it N}^{2}+\pi_{\rm E,\it E}^{2})^{1/2}$ from the parallax parameters 
determined from light-curve modeling. 

In order to estimate the angular Einstein radius, it is needed to convert the measured normalized source radius 
$\rho$ into $\theta_{\rm E}$ by using the angular radius of the source star, i.e. $\theta_{\rm E}=\theta_{\ast}/\rho$. 
The angular source radius is estimated based on the de-reddened color $(V-I)_{0}$ and brightness $I_{0}$ of the 
source star which are calibrated by using the centroid of the Bulge giant clump on the color-magnitude diagram 
as a reference \citep{yoo04}. By adopting the color and brightness of the clump centroid $(V-I,I)_{0,{\rm c}}=
(1.06, 14.45)$ \citep{bensby11,nataf13}, we estimate that $(V-I,I)_{0}=(0.72, 17.68)$ for the source star, 
implying that the source star is a G-type subgiant. Figure~\ref{fig:four} shows the locations of the source and 
the centroid of giant clump in the color-magnitude diagram of stars around the source star. We then translate 
$V-I$ into $V-K$ using the color-color relation of \citet{bessell88} and obtain the angular source radius by 
using the relation between $V-K$ and $\theta_{\ast}$ of \citet{kervella04}. The determined angular source radius 
is $\theta_{\ast}=0.93\pm0.06$ $\mu$as. Then, the Einstein radius is estimated as $\theta_{\rm E}=0.97\pm0.07$ 
mas for the best-fit solution ($u_{0}>0$ model). We note that $u_{0}<0$ model results in a similar Einstein 
radius due to the similarity in the measured values of $\rho$. Combined with the measured Einstein time scale 
$t_{\rm E}$, the relative lens-source proper motion is estimated as $\mu=\theta_{\rm E}/t_{\rm E}=4.90\pm0.35$ 
mas ${\rm yr}^{-1}$. 

With the measured lens parallax and Einstein radius, we determine the mass and the distance to the lens using 
Equation (\ref{eq1}). In Table~\ref{table:two}, we list the physical parameters of the lens system corresponding
to the $u_{0}>0$ and $u_{0}<0$ solutions. We note that the estimated parameters from the two solutions are 
similar. According to the estimated mass, the lens system is composed of a substellar brown-dwarf companion 
and a late-type M-dwarf primary. The distance to the lens is $D_{\rm L}<1.2$ kpc and thus the lens is located 
in the Galactic disk. The projected separation between the lens components is $r_{\perp}=sD_{\rm L}\theta_{\rm E}$ 
is slightly greater than 1 AU. In order to check the validity of the obtained lensing solution, we compute 
the projected kinetic to potential energy ratio $\rm (KE/PE)_{\perp}$ by 
\begin{equation}
\left( {\rm KE \over \rm PE} \right)_{\perp} = {(r_{\perp}/{\rm AU})^{2} \over 8{\pi}^{2}(M/M_{\odot})}
\left[ \left( {1 \over s}{ds \over dt} \right)^{2} + \left( { d\alpha \over dt} \right)^{2} \right]
\label{eq6}
\end{equation}
\citep{dong09}. The estimated ratio is $\rm (KE/PE)_{\perp}<1$ for both $u_{0}>0$ and $u_{0}<0$ solutions and thus 
meets the condition of a bound system.

\begin{deluxetable}{lrr}[h]
\tablecaption{Physical parameters\label{table:two}}
\tablewidth{0pt}
\tablehead{
\multicolumn{1}{c}{Parameter} &
\multicolumn{1}{c}{$u_0>0$} &
\multicolumn{1}{c}{$u_0<0$} 
}
\startdata
Einstein radius (mas)               & 0.97 $\pm$ 0.07   & 0.96 $\pm$ 0.07   \\
Proper motion (mas $\rm yr^{-1}$)   & 4.90 $\pm$ 0.35   & 4.96 $\pm$ 0.35   \\
Total mass ($M_\odot$)              & 0.156 $\pm$ 0.017 & 0.133 $\pm$ 0.011 \\
Mass of primary ($M_\odot$)         & 0.124 $\pm$ 0.014 & 0.107 $\pm$ 0.009 \\
Mass of companion ($M_\odot$)       & 0.032 $\pm$ 0.004 & 0.026 $\pm$ 0.002 \\
Distance (kpc)                      & 1.16 $\pm$ 0.11   & 1.02 $\pm$ 0.08   \\ 
Projected separation (AU)           & 1.16 $\pm$ 0.11   & 1.02 $\pm$ 0.08   \\ 
$({\rm KE}/{\rm PE})_{\perp}$       & 0.48 $\pm$ 0.20   & 0.32 $\pm$ 0.07  
\enddata  
\end{deluxetable}

\section{Conclusion}

We presented the analysis of a caustic-crossing binary-lens microlensing event OGLE-2013-BLG-0578 that led to 
the discovery of a binary system composed of a substellar brown-dwarf companion and a late-type M-dwarf primary. 
Identification of the lens became possible due to the prediction of the caustic crossing from vigilant real-time 
modeling and resolution of the caustic from prompt follow-up observation. The event demonstrates the capability 
of current real-time modeling and the usefulness of microlensing in detecting and characterizing 
faint or dark objects in the Galaxy.

\acknowledgments
Work by C.H. was supported by Creative Research Initiative 
Program (2009-0081561) of National Research Foundation of 
Korea. We acknowledge funding from the European Research 
Council under the European Community's Seventh Framework 
Programme (FP7/2007-2013)/ERC grant agreement No. 246678 
to A.U. for the OGLE project.

\end{document}